# Graphene analogue BCN: femtosecond nonlinear optical susceptibility and hot carrier dynamics

Sunil Kumar,[1,2] N. Kamaraju,[1,2] K. S. Vasu,[1] Angshuman Nag,[3] A. K. Sood,[1-3,*] and C. N. R. Rao[3]

[1]*Department of Physics, Indian Institute of Science, Bangalore 560012, India*

[2]*Center for Ultrafast Laser Applications, Indian Institute of Science, Bangalore 560012, India*

[3]*Chemistry and Physics of Materials Unit and International Center for Materials Science,*

*Jawaharlal Nehru Center for Advanced Scientific Research, Jakkur P.O., Bangalore 560064, India*

[*]*Corresponding author: asood@physics.iisc.ernet.in*

## Abstract

Third-order nonlinear absorption and refraction coefficients of a few-layer boron carbon nitride (BCN) and reduced graphene oxide (RGO) suspensions have been measured at 3.2 eV in the femtosecond regime. Optical limiting behavior is exhibited by BCN as compared to saturable absorption in RGO. Nondegenerate time-resolved differential transmissions from BCN and RGO show different relaxation times. These differences in the optical nonlinearity and carrier dynamics are discussed in the light of semiconducting electronic band structure of BCN vis-à-vis the Dirac linear band structure of graphene.

# 1. Introduction

Graphene, an atomically thick single layer of carbon atoms arranged in a hexagonal lattice, has been of intense interest in recent years, both from the viewpoint of basic physics and for applications in electronics and photonics [1-8]. The remarkably large and wide-band saturable absorption from the visible to the infrared spectral range with ultrafast recovery time in graphene has found applications in wide-band tunable ultrafast mode-locked lasers [9,10]. The ability to manipulate the structure and composition at the nanoscale creates novel materials with superior performance for future technologies. Theoretically, it is possible to build new single layer materials starting from graphene by doping. BCN is an analogue of graphene, prepared and characterized recently [11]. Different electronic band structures of these two-dimensional nano materials lead to their different electronic and optical properties. The electronic band structure calculations using density functional theory [11] show BCN to be a semiconductor with a band gap energy of ~ 1.8 eV, whereas symmetric linear valence and conduction bands at the K points with zero band gap are the hall marks of single-layer graphene. BCN and graphene are two excellent quasi-two-dimensional systems with semiconducting and the semi-metallic properties respectively, which are ideally suited to study the ultrafast dynamics of two-dimensional electron-hole plasma generated after optical excitation and nonlinear optical properties for possible applications in high-speed electronics and photonics. These include ultrafast all-optical switching devices that require a material with not only a fast response but also the one that recovers to its equilibrium state rapidly. Further, optical limiters are important components in photonic applications for device protection from intense radiation. In the femtosecond regime, the nonlinear effects contributing to optical limiting are dominantly electronic rather than thermal as in the nanosecond and picosecond cases. Semiconducting materials, including nanostructures, have been studied in the femtosecond regime [12-16] for photonic applications such as controlling the amplitude, phase and direction of light in all-optical devices.

Transport in steady state measurements is governed by carriers near the Fermi level. In contrast, transport in high speed devices is determined by the dynamic conductivity of hot carriers whose temperature is elevated above the lattice temperature due to presence of high fields and/or rapidly varying fields in the material. It is, therefore, essential to understand the cooling mechanisms of carriers for potential applications of novel materials in high-speed electronic and photonic devices. The ultrafast dynamics of carriers in graphene and very thin graphitic films is decided by the interactions among the carriers and their coupling to the lattice phonons [17-25]. In this article, we present results from measurements of the femtosecond dynamics of photo-excited carriers in a few-layer RGO and BCN.

Our time-resolved experiments utilize a pump (3.2 eV) and probe (1.57 eV) technique with a pulse-duration of $\sim$ 80 fs and examine the time-resolved differential transmission through suspensions as well as thin films of BCN and RGO. Two-component relaxation dynamics ($\tau_1 \sim$ 150 fs and $\tau_2 \sim$ 1 ps) is observed in the case of BCN, whereas a much larger third component ($\tau_3$ > 15 ps) is observed for RGO. Femtosecond nonlinear transmission experiments were performed at 3.2 eV on the BCN and RGO suspensions using the single beam z-scan method to measure the nonlinear absorption and refraction properties. Optical limiting with two photon absorption coefficient, $\beta \sim 1.5 \times 10^{-7}$ cm/W for BCN-suspension is observed, in contrast to saturation absorption for RGO-suspension.

## 2. Experimental Details

Our experiments were performed on BCN-suspensions prepared in iso-propanol and RGO-suspensions prepared in water as well as a few nm thick films deposited on indium tin oxide (ITO) coated glass plates. RGO was prepared by the hydrazine reduction of single layer graphene oxide in water [26]. BCN was prepared by the reaction of high surface area carbon with urea and boric acid at high temperature [11]. The average composition was BCN but there can be regions or domains with somewhat variable compositions. The BCN and graphene concentrations were ~150 μg/ml and 80μg/ml, respectively. To obtain thin films of BCN and RGO, 20 μl of each was drop cast on ITO coated glass plates and allowed to dry for a day. The thickness of each film was ~75 nm.

Nondegenerate time-resolved differential transmission experiments were performed using a standard pump-probe method. A Ti:sapphire laser amplifier (Spitfire, Spectra Physics) provides femtosecond optical pulses centered at 1.57 eV (790 nm) at a pulse repetition rate of 1 kHz. A small fraction of the laser output was used as probe after passing through a variable optical delay line. A thin beta barium borate (BBO) crystal was used to generate second harmonic at 3.2 eV (395 nm) and used as pump to excite the samples. At the sample point, the pulse durations of the probe and the pump pulses were measured to be ~80 fs and 120 fs, respectively. The pump fluence was varied between 60 μJ/cm$^2$ and 590 μJ/cm$^2$, whereas that of the probe was fixed at ~10 μJ/cm$^2$. The probe beam spot size on the sample was ~0.5 mm, very much within the pump spot (~1 mm). The pump beam was modulated at a frequency of 383 Hz using an optical chopper and the change in the transmitted probe beam with respect to a reference was detected using a photodiode as a function of delay between the pump and the probe pulses in a standard lock-in amplifier detection scheme.

Nonlinear transmission experiments were carried out at 3.2 eV using the z-scan method in both open aperture (OA) and closed aperture (CA) configurations. The incident power on the sample was varied by using a focusing lens from ~2 MW/cm$^2$ (far from the focal point) to ~120 GW/cm$^2$ (at the focal point, z = 0). The transmitted power through the sample was measured as a function of the sample position (z) and the change with respect to a reference beam was recorded. In the OA case, all the transmitted intensity was collected using a collection lens, whereas in the CA case an aperture with circular diameter of 3 mm was placed on the photodiode.

## 3. Experimental Results

### 3.1. Nondegenerate time-resolved differential transmission measurements

Figure 1 shows the time-resolved differential transmission spectra of BCN and RGO suspensions and films at the highest pump fluence of 590 μJ/cm$^2$. For sake of clarity, data recorded till 60 ps are shown only till 3 ps. The signal (ΔT/T) remains positive at all probe delays and pump intensities in both the samples. Here, ΔT(t) is the pump-induced change in probe transmission at a time delay t between the pump and the probe pulses, and T is the transmission of the probe in absence of the pump. The initial response is a pulse-width-limited rise in the signal. Thereafter, the signal recovers with a bi-exponential response for the BCN and tri-exponential response for the RGO. The solid lines in Fig. 1 are fit to $\Delta T(t)/T = \left[ H(t) \sum_{i=1}^{m} A_i \exp(-t/\tau_i) \right] \otimes G(t)$. Here the term inside the square brackets is the product of the Heaviside step function $H(t)$ and a sum of exponential functions and the resultant is convoluted with Gaussian cross-correlation function $G(t)$ to account for the laser pulse width. The summation index $m = 2$ for BCN and $m = 3$ for RGO. The amplitudes $A_i$ and time constants $\tau_i$ are given in Tab. 1 for the pump fluence of 590 µJ/cm$^2$. It can be seen from Tab. 1 that $\tau_1 = 200$ fs and $\tau_2 = 1.2$ ps for the BCN-suspension as compared to $\tau_1 = 120$ fs and $\tau_2 = 600$ fs for the BCN-film. On the other hand, the signals from both the RGO-suspension and the RGO-film recover with three time constants, with the largest time constant, $\tau_3$ having the major difference in the two cases.

The dependence of the time constants and $(\Delta T/T)_{max}$ on the pump fluence are presented in Fig. 2 for the RGO-suspension (open circles) and the RGO-film (filled circles) along with those for the BCN-suspension (filled stars). It can be seen that the magnitude of the initial transmission $(\Delta T/T)_{max}$ scales linearly with the pump fluence (Fig. 2a) and the time constants decrease with increasing pump fluence for both the RGO-suspension and the RGO-film (Fig. 2b).

### 3.2. Open and closed aperture z-scan measurements

Nonlinear absorption and refraction coefficients of the samples were measured using a single beam z-scan method. The normalized transmissions in the OA and CA configurations are presented for the BCN-suspension in Fig. 3 and for the RGO-suspension in Fig. 4. The signal from the OA z-scan for BCN (filled circles in Fig. 3a)] shows optical limiting, i.e., reduction in transmission as the input beam intensity is increased, whereas for the RGO (filled circles in Fig. 4a) it shows saturation absorption. The open circles in Fig. 3 and 4 are the results for the corresponding solvents (iso-propanol and water). It can be seen that under the same experimental conditions of OA z-scan, there is no change in the transmission for the solvents. A positive refractive nonlinearity, i.e., decrease in the transmitted intensity due to refraction as the sample approaches the focal point followed by increase in intensity as it moves away from it and towards the detector, can be seen for both the BCN and the RGO suspensions from the CA z-scan (Figs. 3b and 4b). The corresponding results from the solvents alone are presented in the insets of Figs. 3b and 4b.

The intensity-dependent absorption coefficient ($\alpha$) and refractive index ($n$) of a sample can be written as

$$\alpha(I) = \frac{\alpha_0}{1 + I/I_s} + \beta I \tag{1a}$$

$$n(I) = n_0 + \gamma I \tag{1b}$$

Here, $\alpha_0$ is the linear absorption coefficient (single photon absorption), $I_s$ is saturation intensity of single photon absorption, $\beta$ is the two photon absorption (TPA) coefficient and $\gamma$ is the nonlinear refractive index. For saturation absorption behavior, the first term in Eq. 1(a) is important [27] whereas for optical limiting the intensity dependence comes only from the second term. The OA z-scan measurement is sensitive to $\beta$ only whereas CA z-scan signal has contributions both from $\beta$ and $\gamma$ and hence the experimental data are fitted to obtain both of these coefficients simultaneously.

The experimental results for the BCN and RGO suspensions have been fitted numerically [27] and shown by solid lines in Figs. 3 and 4. The values of nonlinear coefficients thus obtained are given in Tab. 2 along with the linear absorption coefficient, $\alpha_0$ measured separately at both 3.2 eV and 1.57 eV for comparison. For the solvents, a value of $\gamma \sim 7 \times 10^{-16}$ cm$^2$/W is obtained for both iso-propanol and water from the numerical fit of the CA z-scan data as shown in the insets of Figs. 3b and 4b, whereas the nonlinear absorption coefficients are negligibly small (OA z-scan data shown by open circles in Figs. 3a and 4a). For the BCN, $\beta \sim 1.5 \times 10^{-7}$ cm/W as compared to

~2.5x10$^{-8}$ cm/W for the RGO, almost an order higher than the RGO. The nonlinear refraction coefficients for RGO and BCN suspensions are similar, $\gamma$ ~5x10$^{-12}$ cm$^2$/W. We note that the measured value of $\beta$ for the RGO-suspension at 3.2 eV is less than half of that at 1.57 eV whereas the saturation intensity $I_s$ is about two times higher [19,20]. Comparing our values of $\beta$ and $\gamma$ for the BCN-suspension at 3.2 eV in the femtosecond regime it is seen that these are higher as compared to other semiconducting films and nanoparticles, for example, for a 5 μm thick semiconducting gallium nitride film: $\beta$ ~3.7x10$^{-8}$ cm/W and $\gamma$ ~-1.5x10$^{-12}$ cm$^2$/W measured at 3.3 eV (370 nm) [13] and for semiconducting cadmium sulfide nanoparticles: $\beta$ ~1.0x10$^{-8}$ cm/W and $\gamma$ ~5.7x10$^{-13}$ cm$^2$/W measured at 1.6 eV (780 nm) [14].

The nonlinear absorption and refraction coefficients can be used to calculate the third order optical susceptibility $\chi^{(3)}$ [28]:

$$\mathrm{Im}\,\chi^{(3)}_{[esu]} = \frac{10^{-7} c \lambda n_0^2}{96 \pi^2} \beta_{[cm/W]} \qquad (2a)$$

$$\mathrm{Re}\,\chi^{(3)}_{[esu]} = \frac{10^{-6} c n_0^2}{480 \pi^2} \gamma_{[cm^2/W]} \qquad (2b)$$

Here $c$ is the speed of light in vacuum, $\lambda$ is the wavelength and $n_0$ is the linear refractive index (taken to be 1.5). These values are given in Tab. 2. For photonic applications such as controlling the amplitude, phase and direction of light of wavelength $\lambda$ in all-optical devices, a figure of merit for the nonlinear material is defined as FOM1 = $\beta\lambda/\gamma$. Another figure of merit that quantifies the two photon absorption process with respect to the single photon absorption is defined as FOM2 = |Im $\chi^{(3)}$|/$\alpha_0$. The measured values of FOM1 and FOM2 for the RGO and the BCN suspensions are given in Tab. 2. It can be seen that both figure of merits are higher for the BCN-suspension as compared to the RGO-suspension.

## 4. Discussion

The mechanisms involved in photo-excitation and relaxation of hot carriers, and optical nonlinearities in BCN and graphene are expected to be different due to the differences in their electronic band structures. The band structure of BCN is that of a direct band gap semiconductor with band gap energy of ~1.8 eV at the $\Gamma$-point in the reciprocal space [11]. The electronic density of states (DOS) of BCN is shown in Fig. 5a (Ref. 11). In Fig. 5b, we show the linear band structure of quasi-two-dimensional semi-metallic graphene near the valence ($\pi$-) and conduction ($\pi$*-) band extrema located at the corners (K- and K′-points) of the hexagonal Brillouin zone along with the electronic density of states: linear function of energy up-to ~2 eV.

Pump excitation with photon energy 3.2 eV generates carriers in the optically coupled states (shown by vertical blue solid arrows in Figs. 5a and 5b). The probe pulse (shown by vertical dashed red arrows in Figs. 5a and 5b) probes the carrier population in the excited state at energy equal to the probe photon energy. In Fig. 5c we show schematically the carrier distribution $f(E)$ at different times after the pump excitation. Immediately after photo-excitation, the carrier thermalization occurs *via* carrier-carrier scattering in time $\tau_1$ to reach a Fermi-Dirac distribution with an elevated electron temperature. The most efficient mechanism for the cooling of carriers is *via* emission of cascade of optical phonons in time $\tau_2$. Further relaxation occurs by conversion of optical phonons into low energy acoustic phonons and/or carrier-acoustic phonon scattering at much longer time scales.

### 4.1. Optical excitation of BCN

Pump-induced photo-bleaching, i.e., positive change in the differential transmission signal ($\Delta$T/T) is observed for BCN (Fig. 1a). The build-up of carrier population in the excited state at the probe photon energy is reflected in the value of $\Delta T/T$ at zero delay between the pump and the probe pulses. Afterwards, the signal decays bi-exponentially and the system reaches the equilibrium ground state within 1.2 ps.

The single photon absorption coefficients of BCN at 1.57 eV and 3.2 eV have been measured by us to be ~9x10$^3$ cm$^{-1}$ and 1.6x10$^4$ cm$^{-1}$, respectively (Tab. 2). The BCN samples do not show a significant $\Delta T/T$ signal from the degenerate pump-probe at 1.57 eV as well as the OA z-scan experiments. This implies that the two-photon absorption coefficient at 1.57 eV is negligibly small within the intensity levels used in our experiments, whereas optical limiting, resulting from two photon absorption processes in BCN-suspension is observed when excited with 3.2 eV photons in the OA z-scan experiments.

Photo-carrier lifetime is defined as the time taken for the transmission change to drop to 1/e of its maximum value. In BCN, the major component is a faster process with time constant $\tau_1$ ~120 fs ($A_1 = 0.97$) (see Tab. 2). In semiconductors, carrier-carrier scattering takes place on a few hundred fs time scale (the carrier mean-free-path in semiconductors is much larger than metals due to the small carrier density at room temperature), followed by the emission of optical phonons for the carriers to reach the band extrema within a few ps. Carrier recombination occurs in a few ps to hundreds of ps depending on the density of mid gap states [12,14,16]. Thus, the faster time constant $\tau_1$ ~120 fs can be associated with carrier-carrier scattering and the slower one, $\tau_2$ ~1 ps to carrier-optical phonon scattering. The Raman spectrum of the BCN film [11] shows a prominent band at ~1590 cm$^{-1}$ having a full width at half maximum of ~60 cm$^{-1}$, corresponding to a lifetime ~600 fs of the optical phonons which is close to $\tau_2$ (~1.2 ps for the BCN-suspension and ~600 fs for the BCN-film) obtained from our time resolved experiments.

BCN can, therefore, be considered a good candidate for an optical switch with a very fast response.

## 4.2. Optical excitation of graphene

Saturation absorption in graphene at a wide range of energies from visible to infrared arises from band filling effects. The pump generated carriers thermalize nearly instantaneously by carrier-carrier scattering within first few tens of fs [25], spreading the carrier distribution over a wide range, both above and below the states optically coupled by the pump. The hot carrier population causes absorption saturation not only at the pump photon energy but also at probe photon energies different from the pump (though to a lesser extent) [29]. The magnitude of saturation absorption and hence the change in probe transmission will be smaller for the nondegenerate case as seen for RGO: $(\Delta T/T)_{max} = 6.2 \times 10^{-3}$ in degenerate pump-probe [20] and $1.9 \times 10^{-3}$ in present nondegenerate pump-probe experiments.

The saturation absorption in graphene recovers with three time constants (Fig. 2b). The carrier-carrier scattering takes place within first few tens of fs [25] that can not be resolved in our experiments. This is followed by emission of optical phonons in a few hundred fs time scale [20,24] *via* intra- and intervalley carrier-phonon scattering, as shown by dotted arrows in Fig. 5b, thus identifying $\tau_1 = 200$ fs for the RGO as carrier-optical phonon scattering time. These optical phonons survive for a few ps before they convert into low energy acoustic phonons via anharmonic decay with relaxation time $\tau_2$ [20,23,24]. The pump-fluence dependence of $\tau_1$ and $\tau_2$ can be seen from Fig. 2b: smaller carrier density at lower pump fluence results in slower cooling time.

The major difference between the carrier dynamics in the RGO-suspension and the RGO-film is the difference in the third time constant. We see that $\tau_3 > 15$ ps for the RGO-suspension and much slower (>150 ps) for the RGO-film. The possible origin of this difference can be due to the interaction between layers in the film, the substrate effect, and different levels of defect density in the two samples. The two possible candidates for $\tau_3$ are carrier-acoustic phonon scattering and interband carrier recombination [21]. Since $\tau_3$ decreases with the increasing pump fluence (Fig. 2b), the interband electron-hole recombination is likely to be responsible for $\tau_3$.

Saturation absorption behavior for the RGO-suspension is seen from OA z-scan measurements (Fig. 4a). The measured value of $\beta \sim 2.5 \times 10^{-8}$ cm/W from the present experiments at 3.2 eV is smaller than that was obtained at 1.57 eV ($\sim 5 \times 10^{-8}$ cm/W) [19,20] whereas the saturation intensity $I_s$ is about two times higher at 3.2 eV. This can arise due to the higher density of states at 3.2 eV (Fig. 5b), thus allowing more number of carriers at 3.2 eV as compared to 1.57 eV, before the absorption is saturated.

## 5. Conclusions

To summarize, we have studied the femtosecond photo-excited carrier dynamics and third-order nonlinear optical susceptibility in a few-layer BCN and RGO at excitation energy of 3.2 eV. Optical limiting behavior is observed in BCN in contrast to saturable absorption in graphene. Nondegenerate pump-probe differential transmission measurements on BCN reveal two-component relaxation dynamics of the photo-generated carriers with time constants ~150 fs and ~1 ps, whereas three time constants ~200 fs, 1.5 ps and >15 ps are observed in the case of RGO. The differences in optical nonlinearity and carrier dynamics have been discussed by considering the semiconducting electronic band structure of BCN compared to the semi-metallic linear band structure of graphene. The figures of merit quantifying the nonlinear absorption with respect to the nonlinear refraction and linear absorption are found to be much better for BCN (Tab. 2) thus making BCN an attractive candidate for ultrafast optical applications.

## Acknowledgements

AKS thanks Department of Science and Technology, India for financial support and SK acknowledges University Grants Commission, India for Senior Research Fellowship.

## References

[1] A. K. Geim and K. S. Novoselov Nat. Mater. 6 (2007) 183.

[2] A. K. Geim, Science 324 (2009) 1530.

[3] C. N. R. Rao, A. K. Sood, R. Voggu and K. S. Subrahmanyam, J. Phys. Chem. Lett. 1 (2010) 572.

[4] C. Berger, Z. Song, X. Li, X. Wu, N. Brown, C. Naud, D. Mayou, T. Li, J. Hass, A. N. Marchenkov, E. H. Conrad, P. N. First, W. A. de Heer, Science 312 (2006) 1191.

[5] P. Avouris, Z. Chen and V. Perebeinos, Nat. Nanotechnol. 2 (2007) 605.

[6] Y.-M. Lin, K. A. Jenkins, A. V. Garcia, J. P. Small, D. B. Farmer and P. Avouris, Nano Lett. 9 (2009) 422.

[7] F. Xia, T. Mueller, Y.-M. Lin, A. V. Garcia, P. Avouris, Nat. Nanotechnol. 4 (2009) 839.

[8] T. Mueller, F. Xia, P. Avouris, Nat. Photonics 4 (2010) 297.

[9] Q. Bao, H. Zhang, Y. Wang, Z. Ni, Y. Yan, Z. X. Shen, K. P. Loh, D. Y. Tang, Adv. Funct. Mater. 19 (2009) 3077.

[10] Z. Sun, T. Hasan, F. Torrisi, D. Popa, G. Privitera, F. Wang, F. Bonaccorso, D. M. Basko and A C. Ferrari, ACS Nano 4 (2010) 803.

[11] K. Raidongia, A. Nag, K. P. S. S. Hembram, U. V. Waghmare, R. Datta and C. N. R. Rao, Chem. Eur. J. 16 (2010) 149.

[12] H. S. Loka, S. D. Benjamin, P. W. E. Smith, IEEE J. Quant. Elect. 8 (1998) 1426.

[13] Y.-L. Huang, C.-K. Sun, J.-C. Liang, S. Kelle, M. P. Mack, U. K. Mishra, S. P. Denbaars, Appl. Phys. Lett. 75 (1999) 3524.

[14] J. He, J. Mi, H. Li, W. Ji, J. Phys. Chem. B 109 (2005) 19184.

[15] H.-M. Gong, X.-H. Wang, Y.-M. Du, Q.-Q. Wang, J. Chem. Phys. 125 (2006) 024707.

[16] R. P. Prasankumar, P. C. Upadhya, A. J. Taylor, Phys. Stat. Solid. B 246 (2009) 1973.

[17] D. Sun, Z.-K. Wu, C. Divin, X. Li, C. Berger, W. A. de Heer, P. N. First, T. B. Norris, Phys. Rev. Lett. 101 (2008) 157402.

[18] J. M. Dawlaty, S. Shivaraman, M. Chandrashekhar, F. Rana, M. G. Spencer, Appl. Phys. Lett. 92 (2008) 042116.

[19] S. Kumar, M. Anija, N. Kamaraju, K. S. Vasu, K. S. Subrahmanyam, A. K. Sood, C. N. R. Rao, Appl. Phys. Lett. 95 (2009) 191911.

[20] S. Kumar, N. Kamaraju, K. S. Vasu, A. K. Sood, Int. J. Nanosci. (to appear 2010).

[21] P. A. George, J. Strait, J. Dawlaty, S. Shivaraman, M. Chandrashekhar, F. Rana, M. G. Spencer, Nano Lett. 8 (2008) 4248.

[22] R. W. Newson, J. Dean, B. Schmidt, H. M. Van Driel, Opt. Exp. 17 (2009) 2326.

[23] T. Kampfrath, L. Perfetti, F. Schapper, C. Frischkorn, M. Wolf, Phys. Rev. Lett. 95 (2005) 187403.

[24] H. Wang, J. H. Strait, P. A. George, S. Shivaraman, V. B. Shields, M. Chandrashekhar, J. Hwang, F. Rana, M. G. Spencer, C. S. R. Vargas, J. Park, Appl. Phys. Lett. 96 (2010) 081917.

[25] M. Breusing, C. Ropers, T. Elsaesser, Phys. Rev. Lett. 102 (2009) 086809.

[26] K. S. Vasu, B. Chakraborty, S. Sampath, A. K. Sood, Solid State Commun. 150 (2010) 1295.

[27] N. Kamaraju, S. Kumar, A. K. Sood, S. Guha, S. Krishnamurthy, C. N. R. Rao, Appl. Phys. Lett. 91 (2007) 251103.

[28] R. A. Ganeev, J. Opt. A: Appl. Opt. 7 (2005) 717.

[29] K. Seibert, G. C. Cho, W. Kütt, H. Kurz, D. H. Reitze, J. I. Dadap, H. Ahn, M. C. Downer, A. M. Malvezzi, Phys. Rev. B 42 (1990) 2842.

**Table captions:**

**Table 1.** Physical parameters obtained from numerical fitting of the time resolved data (normalized) for the BCN and the RGO samples.

**Table 2.** Optical constants of the BCN and RGO-suspension measured using z-scan method at 3.2 eV (395 nm). $\alpha_0^{790}$ is the single photon absorption coefficient measured at 1.57 eV (790 nm).

**Figure captions:**

**Figure 1.** Femtosecond differential probe transmission data for (a) BCN and (b) RGO with 3.2 eV pump and 1.57 eV nm probe pulses. Pump fluence was 590 μJ/cm². Solid lines are fits of the data to a bi-exponential (tri-exponential) response function for BCN (RGO) convoluted with Gaussian pump-and-probe temporal profiles. The inset of (b) shows the signals (up to 50 ps) from the RGO-suspension and the RGO-film in log-log scale clearly showing three distinct regions characterized by three time constants $\tau_1$, $\tau_2$ and $\tau_3$ of the carrier dynamics (see text).

**Figure 2.** Summary of the results from nondegenerate pump-probe experiments on the RGO-suspension (open circles), RGO-film (filled circles) and BCN-suspension (filled stars) as a function of pump fluence. The solid lines in (a) are linear fits.

**Figure 3.** Z-scan results for the BCN-suspension (filled circles) and iso-propanol (open circles) in the (a) OA, and (b) CA configurations. The continuous lines are numerical fits.

**Figure 4.** Z-scan results for the RGO-suspension (filled circles) and water (open circles) in the (a) OA, and (b) CA configurations. The continuous lines are numerical fits.

**Figure 5.** (a) Electronic density of states in BCN (Ref. 11), (b) Schematic electronic band structure and density of states near the Dirac points in graphene. The pump excitation at 3.2 eV is indicated by vertical blue arrows and the dashed red arrows represent the transitions by the probe. (c) Nonequilibrium carriers (electrons and holes) distribution at pump excitation and distributions after thermalization$(\tau_1)$ and relaxation *via* intra- and intervalley carrier-phonon scattering ($\tau_2$).

## Table 1

| Sample | ΔT/T $[10^{-4}]$ | $A_1$ | $A_2$ | $A_3$ | $\tau_1$ [ps] | $\tau_2$ [ps] | $\tau_3$ [ps] |
|---|---|---|---|---|---|---|---|
| BCN-film | 8 | 0.97 | 0.03 | - | 0.12 | 0.6 | - |
| BCN-suspension | 13 | 0.97 | 0.03 | - | 0.20 | 1.2 | - |
| RGO-film | 28 (22) | 0.83 (0.88) | 0.12 (0.09) | 0.05 (0.03) | 0.20 (0.17) | 1.7 (1.1) | 150 (25) |
| RGO-suspension | 18.6 (62) | 0.86 (0.88) | 0.11 (0.09) | 0.03 (0.03) | 0.20 (0.19) | 1.6 (1.5) | 15 (25) |

The results in parentheses are from degenerate pump-probe measurements [Ref. 20] on the RGO samples as used in the present experiments and at similar pump fluence of ~ 600 μJ/cm².

## Table 2

| | $\alpha_0$ $[10^4\,cm^{-1}]$ | $\beta$ [cm/GW] | $\gamma$ $[cm^2/GW]$ | $I_s$ $[GW/cm^2]$ | Re$\chi^{(3)}$ [esu] | Im$\chi^{(3)}$ [esu] | FOM1 | FOM2 [esu.cm] | $\alpha_0^{790}$ $[10^4\,cm^{-1}]$ |
|---|---|---|---|---|---|---|---|---|---|
| RGO | 1.9 | 25 | $7.5\times10^{-3}$ | 170 | $1\times10^{-10}$ | $7\times10^{-12}$ | 0.13 | $3.6\times10^{-16}$ | 1.4 |
| BCN | 1.6 | 150 | $5.0\times10^{-3}$ | - | $7\times10^{-11}$ | $4\times10^{-11}$ | 1.18 | $2.4\times10^{-15}$ | 0.9 |

Sunil Kumar et al.

Figure 1

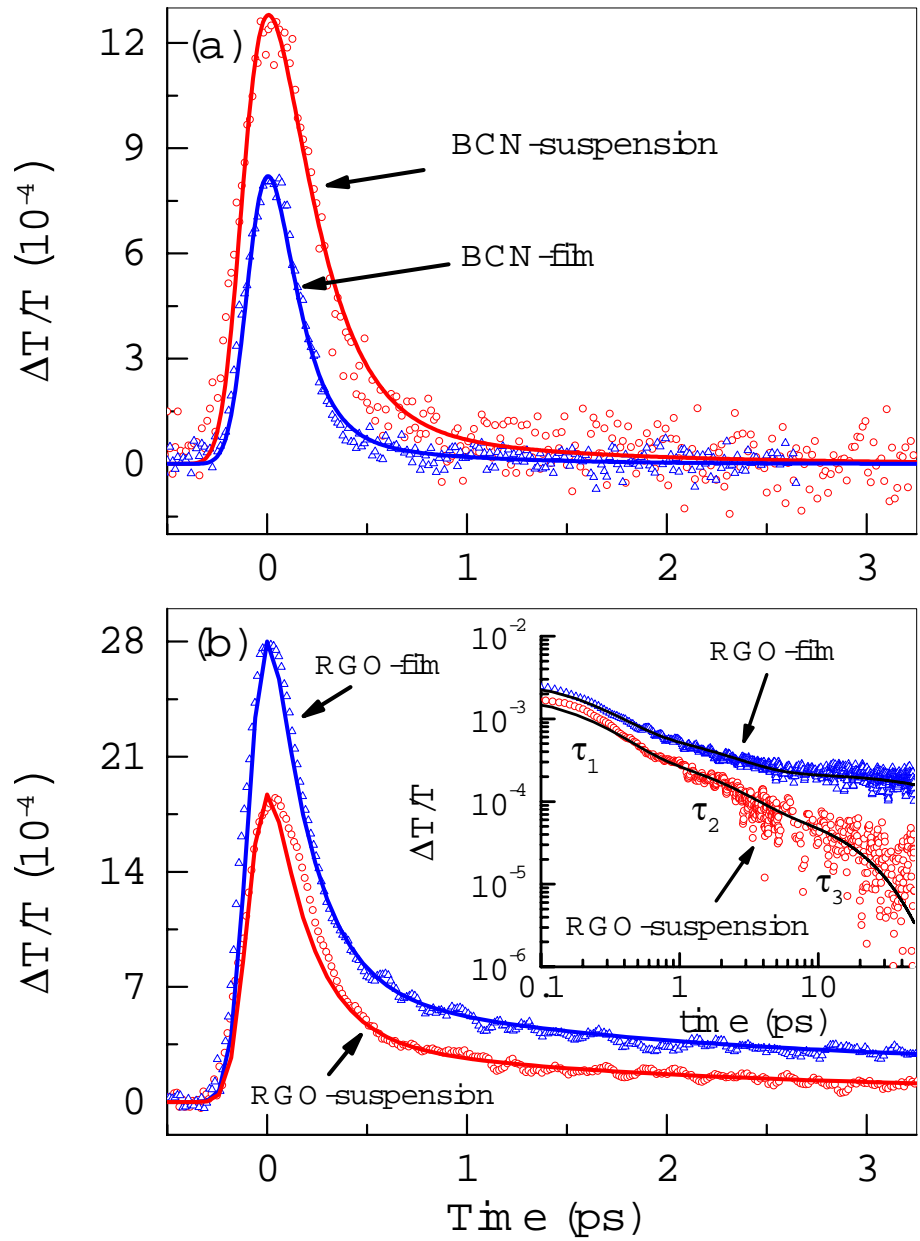

Figure 2

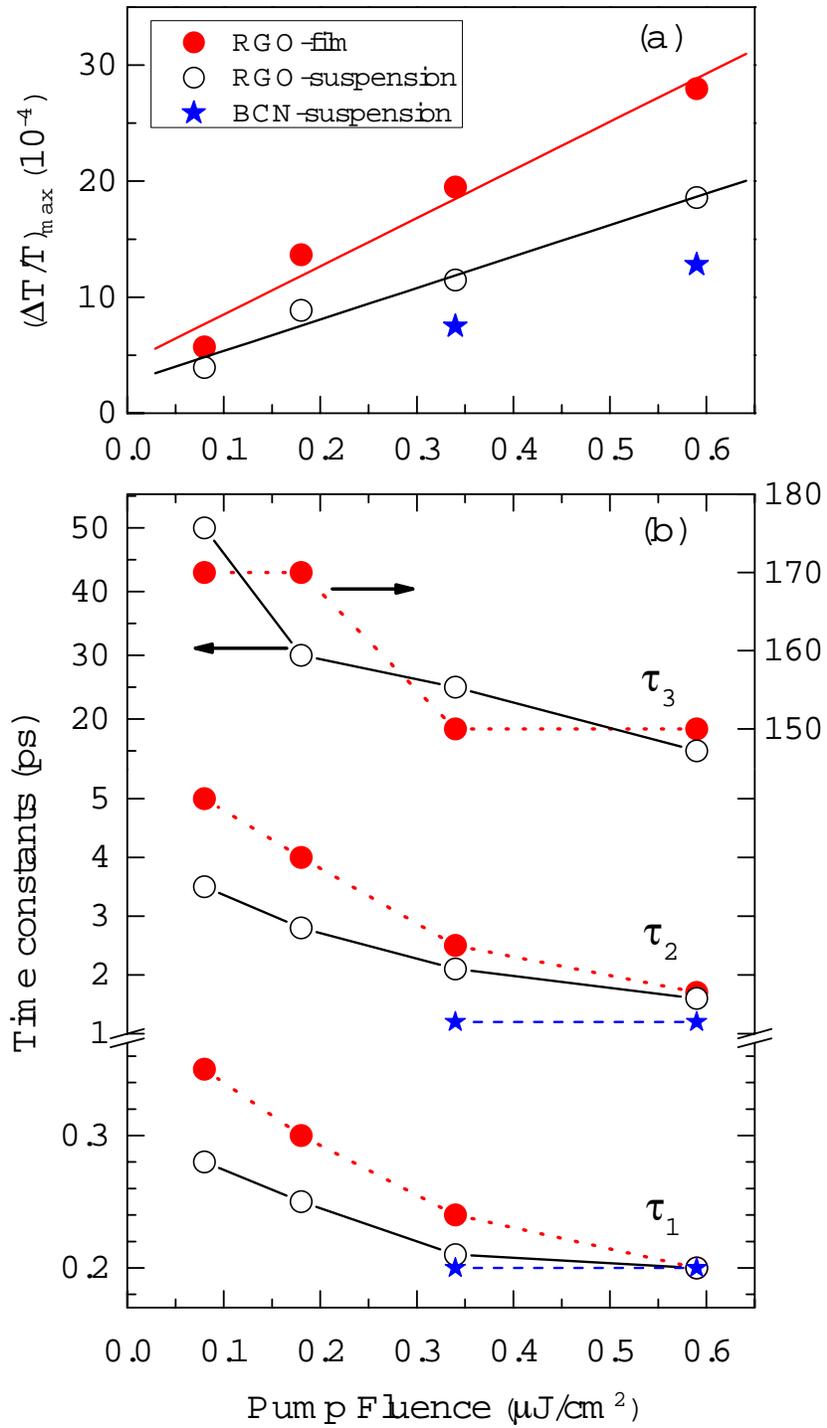

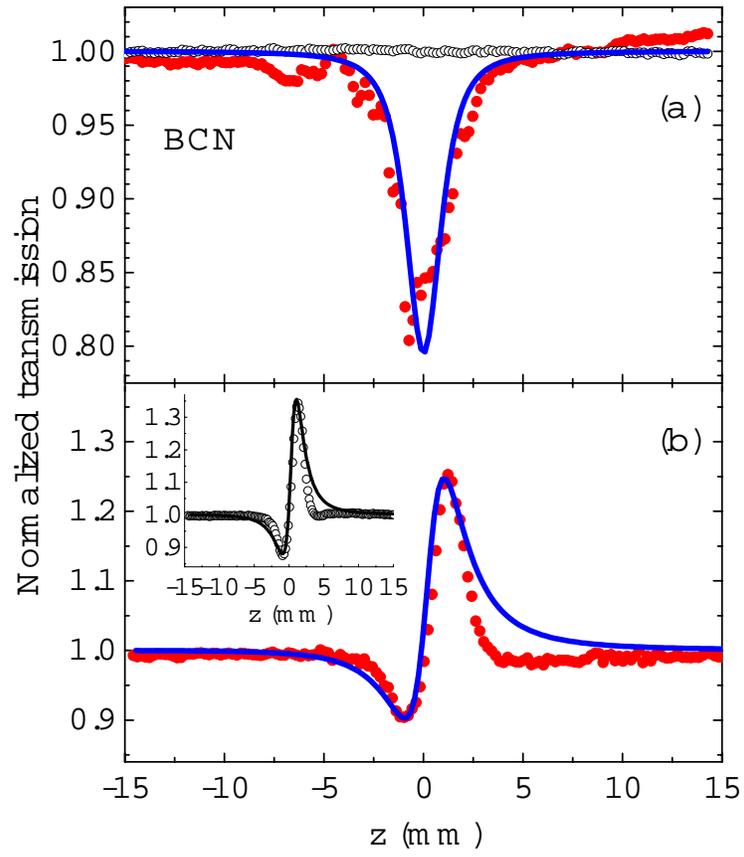

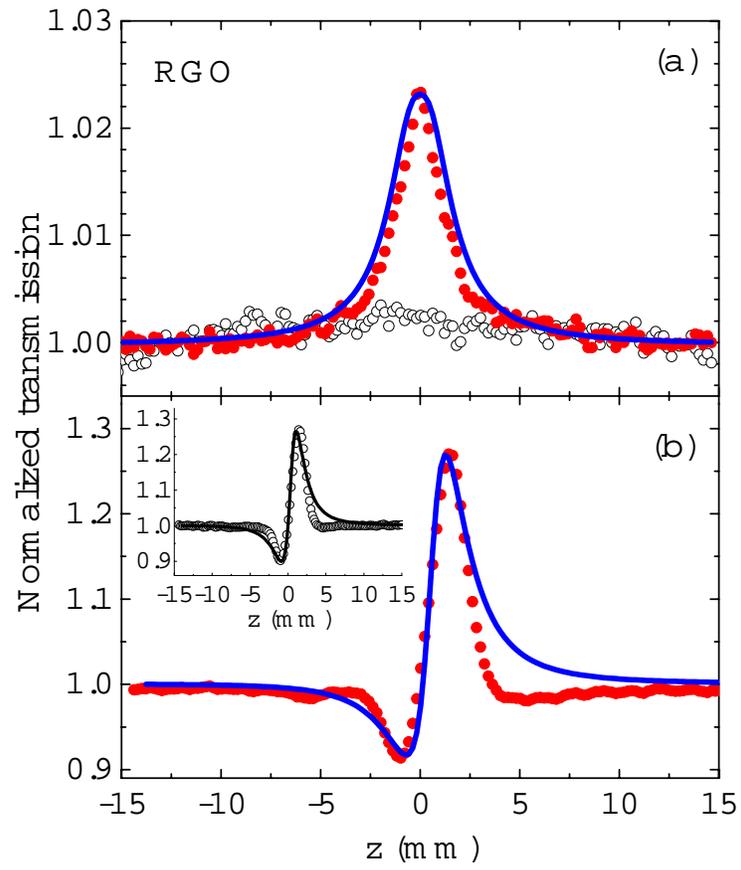

Figure 5

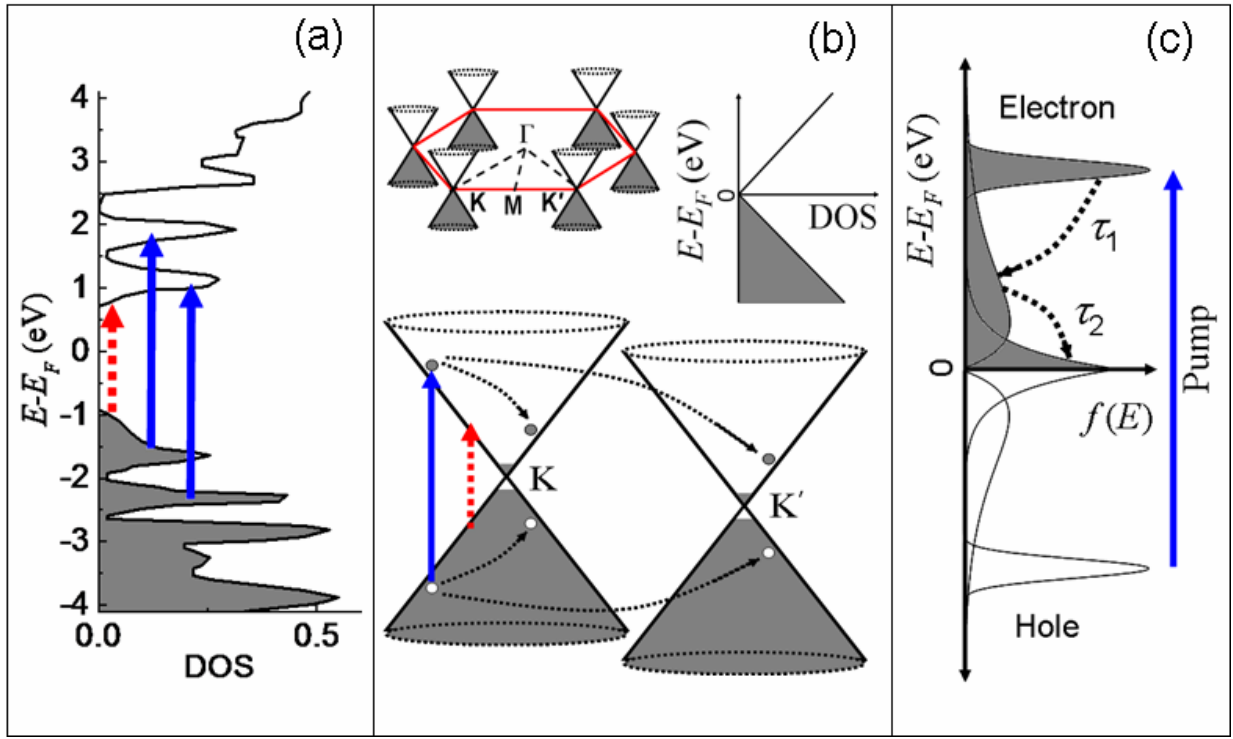